\documentclass[preprintnumbers,nofootinbib,noshowpacs,eqsecnum,prd]{revtex4}

\usepackage{graphicx,epsfig}
\usepackage{amsmath,amssymb,bbold}
\usepackage{url}
\usepackage{rotating}
\usepackage{axodraw}
\usepackage{color}
\usepackage{hhline}

\graphicspath{ {Plots/} }

\makeatletter       

\renewcommand{\p@subsection}{}

\makeatother

\begin{document}

\title{Theoretical Basis of Higgs-Spin Analysis in \boldmath{$H \to
       \gamma\gamma$} and \boldmath{$Z\gamma$} Decays }

\author{S.Y. Choi$^{1}$, M.M. Muhlleitner$^{2}$, and P.M. Zerwas$^{3}$\\[-3mm]
        \mbox{ }\\
   $^1$ {\it Department of Physics, Chonbuk National University, Jeonbuk 561-756,
        Korea}  \\
   $^2$ {\it Institute of Theoretical Physics, Karlsruhe Institute of Technology,
         D-76128 Karlsruhe, Germany} \\
   $^3$ {\it Deutsches Elektronen-Synchrotron DESY, D-22603 Hamburg, Germany}}

\date{\today}

\begin{abstract}
{\it \noindent
     We chart the theoretical basis of radiative decays of the Higgs boson,
     $H \to \gamma\gamma$ and $Z\gamma$, for measuring the spin of the Higgs
     particle. These decay channels are complementary to other rare modes
     such as real/virtual $Z$-boson pair-decays. In systematic helicity analyses
     the angular distribution for zero-spin is confronted with hypothetical
     spin-$2^\pm$ and higher assignments to quantify the sensitivity.}
\end{abstract}

\maketitle

\addtocounter{section}{1}

\noindent
{\bf 1.} After the discovery of the Higgs particle, the properties must be
examined experimentally to identify the particle as the element proper of the
Higgs mechanism for breaking the electroweak symmetries \cite{Higgs}
[for recent general reviews see Refs.$\,$\cite{HiggsR}]. The dynamical steps
include the confirmation of the scalar spin-zero character of the
particle\footnote{This letter expands part of a talk
delivered by S.Y.~Choi at the APCTP 2012 LHC Physics Workshop at Korea,
Seoul, Korea, August 7-9, 2012.}.
Higgs-boson decays to pairs of $Z$-bosons{\footnote{In continuum final states
of Higgs-boson decays $H \to Z+[f\bar{f}]$ the fermion pair $[f\bar{f}]$ is not
necessarily decay product of a virtual $Z^\ast$ state so that general
experimental analyses are not constrained by Bose symmetry.}},
cf. Refs.$\,$\cite{Hspin1,GaoDeR,Hspin2} and additional references listed
there, are promising candidates for examining the Higgs spin at the LHC,
supplemented by other measurements in the Higgs-strahlung processes at LHC
\cite{Hspin22} and  $e^+e^-$ linear colliders \cite{Hspin3}. Exploiting the
primary Higgs-boson search channels, $\gamma\gamma$ and $WW$ decays provide
another self-evident tool for spin measurements \cite{Hspin4,Alves}.
[For a variety of methods see the recent literature in Refs.$\,$\cite{varmeth}].
All these channels are particularly difficult to analyze for low masses of the
Higgs boson where the $b$-pair decays are overwhelming. However, with about
126 GeV this is precisely the mass range, where a new boson has been discovered
by the LHC experiments \cite{LHC_experiments1,LHC_experiments2,LHC_experiments},
consistent with analyses of electroweak precision data \cite{Erler} and
compatible with Higgs patterns within the framework of the Standard Model
\cite{Englert}. \\

In this study we examine to which extent radiative decays \cite{Hgg,HZG,HZgphen}
\begin{eqnarray}
   H \to \gamma\gamma \;\, {\rm and} \;\, Z \gamma
\label{eq:Hgamma}
\end{eqnarray}
can be exploited to determine the spin of the Higgs boson. Even though these
decays are rare, with $BR \simeq 2 \cdot 10^{-3}$ for masses of the order of
126~GeV~\cite{HZgcode,Mond,Djou}, the large number of light Higgs bosons
produced in gluon fusion \cite{Mond,Djou,ggfusH} generates a sample
of order $10^4$ events in $\gamma\gamma$ decays for an LHC luminosity of
$\sim 100$ fb$^{-1}$, and a similar number of $Z\gamma$-decay events
[though reduced finally by the branching ratio for leptonic $e,\mu$ decays
of the $Z$-boson]. These channels nicely supplement the other rare
$ZZ^\ast$ channels. In the present report we perform general helicity
analyses of the radiative decays to quantify the sensitivity to zero-spin
of the Higgs boson in angular distributions. \\[2mm]

\begin{figure}[ht]
\begin{center}
\epsfig{file=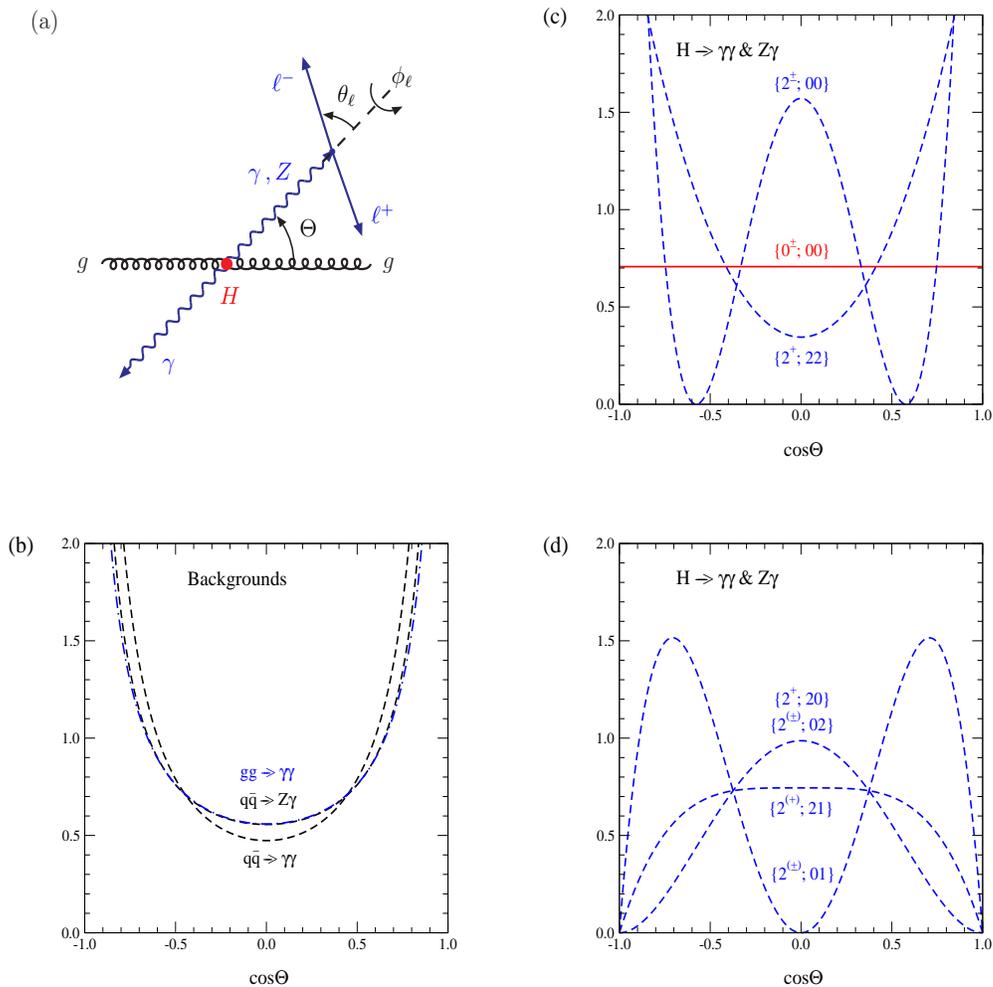,height=6.5cm,width=6.cm}
\hskip 1.cm
\epsfig{file=polar_dist_rr_Zr_set1.eps,height=6.cm,width=6.cm}\\[10mm]
\epsfig{file=polar_dist_bkg.eps,height=6.cm,width=6.cm}
\hskip 1.cm
\epsfig{file=polar_dist_rr_Zr_set2.eps,height=6.cm,width=6.cm}
\end{center}
\caption{\it (a) Kinematics of radiative Higgs-boson decays in gluon fusion;
         (b-d) Angular distributions of $\gamma\gamma$ and $Z\gamma$ axes
         in the rest frame of the subprocesses: the flat Higgs signal compared
         with the distributions of the background events with the invariant
         $q\bar{q}$ energy set to $M_H=126$ GeV in (b) and all potential spin-2
         distributions corresponding
         to the spin component along the collider
         axis in the rest frame of the Higgs boson,
         and the difference of the helicities of the vector bosons in the decay
         final states [all distributions normalized over the interval
         $|\cos\Theta| \leq 1/\sqrt{2}$]. The upper indices refer to the allowed
         parity associated with the distributions in $\gamma\gamma$ decays.
         Allowed parities of $Z\gamma$ states are denoted in round brackets
         either if unique for $Z\gamma$ or if different from $\gamma\gamma$
         final states. [Note that $Z\gamma$ states do not conform with the
         $gg$ initial states in contrast to $\gamma\gamma$ states.]
         }
\label{fig:Fangle}
\end{figure}

\noindent
{\bf 2.} The zero-spin character of the Higgs boson reflects itself in the
isotropic decay distribution of the radiative decays in the rest frame,
i.e.
\begin{equation}
\frac{1}{\Gamma_{\gamma\gamma,Z\gamma}}\,
\frac{d\Gamma_{\gamma\gamma,Z\gamma}}{d\cos\Theta}
= 1  \;\,{\rm and}\;\, 1/2
\label{eq:SM_Higgs_distribution}
\end{equation}
with $\Theta$ denoting the polar angle of the $\gamma\gamma$ and $Z\gamma$ axis,
defined, for example, in the Higgs-boson rest frame with regard to the LHC beam
axis, Fig.$\,${\ref{fig:Fangle}}(a), and with the normalization including the
proper statistical factor of the two-photon state. The axis can be reconstructed
experimentally in the exclusive $\gamma\gamma$ and $Z\gamma$ final states.
Since the $Z$-boson is produced in helicity $\pm 1$ states, the $\ell\ell$
distribution in leptonic $Z$ decays is of the familiar $1+\cos^2\theta_\ell$ form,
$\theta_\ell$ being the polar lepton angle in the rest frame of the $Z$-boson. \\

The flat distribution (\ref{eq:SM_Higgs_distribution}) is unique to zero-spin
of the Higgs boson. For any other spin assignment $J$ the distribution would be
given by the Wigner functions
$\sim |d^J_{m,\lambda_{\gamma} - \lambda^\prime_\gamma}(\Theta)|^2$
and $|d^J_{m,\lambda_{Z} - \lambda^\prime_\gamma}(\Theta)|^2$, where $m$ denotes
the $S_z$ spin component being either 0 or $\pm 2$ for Higgs-boson production in
gluon fusion, while $\lambda_\gamma -\lambda^\prime_\gamma= 0, \pm 2$
and $\lambda_Z - \lambda^\prime_\gamma = 0, \pm 1, \pm 2$.
The explicit form of the Wigner $d$-functions $d^J_{m\lambda}(\Theta)$ may be
read off the tables in Ref.$\,$\cite{Wigner}. \\

The general parity-invariant polar angular distribution can be cast
into a compact form for any Higgs-spin $J$ decays to $\gamma\gamma$
and $Z\gamma$ in gluon fusion:
\begin{eqnarray}
\frac{1}{\sigma} \, \frac{d\sigma [\gamma\gamma]}{d\cos\Theta}
   \,\,\,= \,\,\,
   && \mbox{ }\hskip -0.5cm  (2J+1)\, \left[
     {\mathcal{X}}^J_0 {\mathcal{Y}}^J_0 \,{\mathcal{D}}^J_{00}
   +        {\mathcal{X}}^J_0 {\mathcal{Y}}^J_2 {\mathcal{D}}^J_{02}
          + {\mathcal{X}}^J_2 {\mathcal{Y}}^J_0 {\mathcal{D}}^J_{20}
          + {\mathcal{X}}^J_2 {\mathcal{Y}}^J_2 \,{\mathcal{D}}^J_{22}
   \right]
   \label{eq:Wigner_rr}
\end{eqnarray}
and
\begin{eqnarray}
\frac{1}{\sigma} \, \frac{d^2\sigma [Z\gamma]}{d\cos\Theta\, d\cos\theta_\ell}
   \,\,\,= \,\,\,
   && \mbox{ } \hskip -0.6cm \frac{3(2J+1)}{16}\,
     \bigg\{ \left[ {\mathcal{X}}^J_0 {\mathcal{Y}'}^J_0 \,{\mathcal{D}}^J_{00}
   +       {\mathcal{X}}^J_0 {\mathcal{Y}'}^J_2 {\mathcal{D}}^J_{02}
   +       {\mathcal{X}}^J_2 {\mathcal{Y}'}^J_0 {\mathcal{D}}^J_{20}
   + {\mathcal{X}}^J_2 {\mathcal{Y}'}^J_2 \,{\mathcal{D}}^J_{22}
      \right] \, (1+\cos^2\theta_\ell)                      \nonumber\\
 && {\hspace{14mm}}
   + 2 \left[ {\mathcal{X}}^J_0 {\mathcal{Y}'}^J_1 \,{\mathcal{D}}^J_{01}
            + {\mathcal{X}}^J_2 {\mathcal{Y}'}^J_1\,
              {\mathcal{D}}^J_{21} \right]\,
            \sin^2\theta_\ell\,\bigg\}
\label{eq:Wigner_Zr}
\end{eqnarray}
respectively, with the squared Wigner functions
${\mathcal{D}}^J_{m \lambda} = \frac{1}{4}\, \sum [{d}^J_{m \lambda}(\Theta)]^2$
symmetrized by the summing over all signs $\pm m, \pm \lambda$ of the
spin/helicity components [index 0 counting twice]. The functions read
explicitly for $J = 0$ and $J = 2$:
\begin{eqnarray}
&& \mathcal{D}^0_{00} = 1   \qquad\qquad\qquad\qquad\qquad\qquad\quad\;\;\;\,
   \mathcal{D}^2_{20} = \mathcal{D}^2_{02} = 3 \sin^4 \Theta /8                          \nonumber\\
&& \mathcal{D}^2_{00} = (3\cos^2\Theta - 1)^2 /4   \qquad\qquad\qquad\quad\;
   \mathcal{D}^2_{01} = \mathcal{D}^2_{10} = 3 \sin^2 \Theta \cos^2 \Theta /2            \nonumber\\
&& \mathcal{D}^2_{22} = (\cos^4\Theta + 6 \cos^2\Theta + 1) / 16  \qquad\quad\,
   \mathcal{D}^2_{21} = \mathcal{D}^2_{12} = \sin^2 \Theta (1+\cos^2\Theta)/4    \,,
\label{angles}
\end{eqnarray}
the functional forms displayed in Fig.$\,${\ref{fig:Fangle}}(c,d).
The [non-negative] reduced production and decay helicity probabilities
$\mathcal{X}$ for gluon fusion, $\mathcal{Y}$ for $\gamma\gamma$ and
$\mathcal{Y}'$ for $Z\gamma$ final states, are model-dependent parameters,
obeying the sum rules
\begin{eqnarray}
\mathcal{X}^J_0 + \mathcal{X}^J_2 &=& 1                                \\
\mathcal{Y}^J_0 + \mathcal{Y}^J_2 &=& 1 \quad {\rm and}\quad
\mathcal{Y}^{\prime J}_0 + \mathcal{Y}^{\prime J}_1
 + \mathcal{Y}^{\prime J}_2 = 1                                        \,.
\end{eqnarray}
[The formalism can easily be transferred to $q\bar{q}$ production with
$S_z = 0$ or $\pm 1$ by substituting $\mathcal{X}^J_1$ for
$\mathcal{X}^J_2$
and ${\mathcal{D}}^J_{1k}$ for ${\mathcal{D}}^J_{2k}$ $(k=0,1,2)$ in the cross
sections and in the $\mathcal{X}$ sum rule. Note however that the rate for signal
Higgs production in $q\bar{q}$ collisions is negligibly small for light
quark beams. The initial states $gg$ and $q \bar{q}$ mix incoherently in the
most general configurations.] \\

The polar angle of the $Z$ decay distribution can easily be integrated out,
equivalent to the substitutions $(1+\cos^2\theta_\ell) \rightarrow 8/3$ and
$\sin^2\theta_\ell \rightarrow 4/3$.  Since the reduced helicity probabilities are
non-negative, the coefficients of both the $(1+\cos^2\theta_\ell)$ term
and the $\sin^2\theta_\ell$ term inevitably generate non-vanishing maximum/minimum
$\pm \cos^{2J}\!\Theta$ terms. Thus, observing [in the experimentally ideal
case] that the angular distribution is independent of $\cos\Theta$, proves
unambiguously the spin-zero character of the Higgs particle. At the same time
the $\sin^2\theta_\ell$ term in $Z\gamma$ final states is predicted to be absent,
providing an independent cross-check. \\

\begin{center}
\begin{table} [htb]
\begin{tabular}{|c||c|c|c|c|}
\hline
$\;{\mathcal{P}} \;\, \backslash \; J\;$  &  $\; 0\;$  & $\; 1 \;$
                                       & $\;$ $2, 4, \cdots$
                                       & $\;$ $3, 5, \cdots$             \\
\hline\hline
$\;$ even $\;$       &      1
                     & $\;$ forbidden $\;$
                     & $\; {\mathcal{D}}^J_{00} \;\; {\mathcal{D}}^J_{02} \;$
                     & $\; {\mathcal{D}}^J_{22}$     \\
                     &
                     &
                     & $\; {\mathcal{D}}^J_{20} \;\; {\mathcal{D}}^J_{22} \;$
                     &                               \\
\hline
$\;$ odd $\;$        &      1
                     & $\;$ forbidden $\;$
                     & $\; {\mathcal{D}}^J_{00}$
                     & $\;$ forbidden $\;$           \\
\hline
\end{tabular}
\caption{\it{Selection rules for Higgs parity following from observing the
             polar angular distribution of a spin-$J$ Higgs state in the process
             $gg\to H\to \gamma\gamma$.}}
\label{tab:SR}
\end{table}
\end{center}

The observation of spin states in $gg \to H \to \gamma\gamma$ allows [partial]
conclusions also on the parity of the states. From Bose symmetry and parity
symmetry of the helicity amplitudes
${\mathcal{T}}^J_{\lambda_1\, \lambda_2}
 = (-1)^J {\mathcal{T}}^J_{\lambda_2\, \lambda_1}
 = {\mathcal{P}} (-1)^J {\mathcal{T}}^J_{-\lambda_1\, -\lambda_2}$,
referring separately to initial and final states \cite{Hspin1},
the selection rules presented in Tab.$\,${\ref{tab:SR}} can easily be
derived [see also Ref.$\,$\cite{Yang}], complementing global rules noted
earlier in the literature. Scalar and pseudoscalar Higgs bosons
cannot be discriminated in the $\gamma\gamma$ decay mode, neither even/odd
parity by observing ${\mathcal{D}}^J_{00}$, and spin correlation
effects \cite{Hspin3,Coleppa:2012eh,PRZ} must be exploited for discrimination.
But observing any state with helicity difference $\Delta\lambda = 2$ in
initial or final state determines unambiguously the even-parity character
of the Higgs boson. Analogous rules apply also to $Z\gamma$ final states in
gluon fusion and $\gamma\gamma$ final states in $q\bar{q}$ annihilation.
Either the first or the second index $\Delta\lambda$ in the $\mathcal{D}$
functions coming with the production or decay amplitude, respectively, is
restricted in parallel to the rules in Tab.$\,${\ref{tab:SR}} while the
respective companion index is unrestricted apart from the standard spin
constraints.  \\

The $\gamma\gamma$ channel is described by two-by-two independent probabilities
for production and decay, ${\mathcal{X}}^J_{0,2}$ and ${\mathcal{Y}}^J_{0,2}$.
Popular choices for experimental simulations are the complementary $\{J;00\}$
$'$scalar-type$'$ and the $\{J;22\}$ $'$tensor-type assignments \!$'$:
\begin{eqnarray}
{\rm 'scalar\mbox{-}type \;\, assignment'}\;&:&
    {\mathcal{X}}^J_0 = {\mathcal{Y}}^J_0 =1 \;\,{\rm and}\;\,
    {\mathcal{X}}^J_2 = {\mathcal{Y}}^J_2 = 0 \quad [J \geq 0]
    \hskip 3.1cm       \\
{\rm 'tensor\mbox{-}type \;\, assignment'}\;&:&
    {\mathcal{X}}^J_0 = {\mathcal{Y}}^J_0 =0 \;\,{\rm and}\;\,
    {\mathcal{X}}^J_2 = {\mathcal{Y}}^J_2 = 1 \quad [J \geq 2]       \,,
\end{eqnarray}
supplemented by the
\begin{eqnarray}
\hspace{10mm} {\rm 'mixed\mbox{-}type \;\, assignment'}\;&:&
    {\mathcal{X}}^J_0 = {\mathcal{Y}}^J_2 =0 \;\,{\rm and}\;\,
    {\mathcal{X}}^J_2 = {\mathcal{Y}}^J_0 = 1 \quad [J \geq 2]
     \quad \mbox{and $'$1,0$\,'$ interchanged}\,,
\end{eqnarray}
for the signal $J=0$ and the hypothetical alternatives $J \geq 2$.
[The $\gamma\gamma$ coupling of the tensor-type assignment
is equivalent to the KK graviton coupling in $d=5$ scenarios \cite{Hspin4}.]
The configurations can be exploited in two ways:
{\it (i)} One of the two scalar- or tensor-type configurations
for $J \geq 2$ for instance, is sufficient to prove that the
spin-zero test of the signal is non-trivial; {\it (ii)} However, to prove
experimentally that the spin-2 assignment is not realized, the three
configurations, which are mutually independent, must necessarily be shown
absent. Not observing the double index $\{J;00 \}$, the state $\{J^- \}$
is ruled out, while neither observing distributions carrying at least one
index 2, the state $\{ J^+ \}$ is ruled out, too.
Thus any spin $J \geq 2$ can be rejected for both parities $\pm$ by angular
analyses. -- Due to potential longitudinal $Z$ polarization accounted for by
${{\mathcal{Y}}'}^J_1$, the $Z\gamma$ final state is described by three
independent decay probabilities. \\

Disregarding $J=1$, as forbidden by the Landau-Yang theorem in $\gamma\gamma$
decays \cite{Landau,Yang}, we will choose $J=2$ for illustration.
[Part of the distributions have also been noted in Refs.~\cite{Hspin4,G1}.]
None of the possible helicity states would generate a flat distribution like
spin-zero Higgs-boson decays. It is shown in Fig.$\,${\ref{fig:Fangle}} how
the flat signal distribution contrasts with the distributions generated
by hypothetical spin $J=2$ assignments of both even and odd parity,
and the angular distribution of the backgrounds as well [to be analyzed next].
In the final section we will compare the spin-0 distribution, assigned $J=0$ and
$m=\lambda_{\gamma,Z}-\lambda^\prime_\gamma=0$, specifically with the
scalar-type $\{2;00\}$ and the tensor-type $\{2;22\}$ distributions,
representing the state $J = 2$ with the two parities $\pm$.
The $\{ 2;00 \}$ scalar-type distribution is pronounced in the center
like spin-0 after angular cut, and rises in the forward/backward directions
like the continuum backgrounds, thus providing a non-trivial analogue
to be discriminated experimentally from the Higgs signal. The  $\{ 2;22 \}$
tensor-type distribution rises monotonically to the left and to the right
of the center, providing also a valuable discriminant. Both types must be ruled
out necessarily to reject experimentally the spin-2 assignment for even and
odd parity. \\

\begin{table}[hbt]
\begin{tabular}{|l||r|cc||l|}
\hline
 & \;\; Polar Moments $\langle |\cos\Theta| \rangle/\langle 1\rangle$\;\;
 & \;\;\; $total$ \;\;\;
 & \;\;\; $cut$  \;\;\;
 & \; $Z$ Decay  \\
\hline\hline
\;$\gamma\gamma, Z\gamma$\;  & spin-0 $\{0;00\}$ {\hspace{15mm}}   & 1/2
                         & $\sqrt{2}/4 \;$
                         &\; 1 \;                                       \\
\hline\hline
\;$\gamma\gamma, Z\gamma$\; & spin-2 $\{2;00\}$ {\hspace{15mm}}  & 5/8
                         & $5\sqrt{2}/36 \;$
                         &\; $1+\cos^2\theta_\ell$\;                   \\
                         &        $\{2;02\}$ {\hspace{15mm}}    & 5/16
                         & $35\sqrt{2}/172 \; $                   &        \\
                         &        $\{2;20\}$ {\hspace{15mm}}    & 5/16
                         & $35\sqrt{2}/172  \;$                   &        \\
                         &        $\{2;22\}$ {\hspace{15mm}}    & 65/96
                         & $155\sqrt{2}/492 \;$
                         &                                           \\
\hline
\; $Z\gamma$\;           & spin-2 $\{2;01\}$ {\hspace{15mm}}   & 5/8
                         & $5\sqrt{2}/14 \;$
                         &\; $\sin^2\theta_\ell$\;                          \\
                         &        $\{2;21\}$ {\hspace{15mm}}   & 5/12
                         & $55\sqrt{2}/228 \;$
                         &                                           \\
\hline
\end{tabular}
\caption{\it Ratios of the first over zeroth moments of the polar
angular distributions of the $\gamma\gamma$ and $Z\gamma$ axes, $|\cos\Theta|$,
for zero Higgs spin and for spin 2 of a hypothetical resonance.
$'$cut $'$ denotes the theoretical cut $|cos\Theta| < 1/\sqrt{2}$ on the
polar angle of the event axes.}
\label{tab:mom}
\end{table}

A first global comparison between spin-0 and all the spin-2 distributions
is offered by the moments of the polar-angle distributions of the $\gamma\gamma$
and $Z\gamma$ event axes{\footnote{Angular asymmetries \cite{Alves}
can equivalently be used in global analyses.}} noted in Tab.$\,${\ref{tab:mom}}.
The first moments of the spin-2 distributions are characteristically
different{\footnote{The assignments $\{2;01\}$ and $\{2;21\}$
can also be discriminated from $\{0;00\}$ by identifying the $Z$-decay angular
distributions $\sin^2\theta_\ell$ versus $(1+\cos^2\theta_\ell)$.}} from the
spin-0 distribution and may provide early information on the Higgs spin.\\

\noindent
{\bf 3.} The large continuum background generated in $pp (q \bar{q}) \to
\gamma\gamma$ and $Z\gamma$ processes [and, to a lesser extent, loop-induced
gluon fusion] requires stringent cuts in order not to dwarf the signals.
The angular characteristics allow to reduce the continuum backgrounds
considerably. \\

The angular distribution of the background events is strongly peaked in the
forward and backward directions as a result of the $t$ and $u$-channel
exchange mechanisms, in contrast to the flat distribution of the signal. In the
rest frame of the parton system, cutting out the singular forward and backward directions
$\Theta \to 0$ and $\pi$ [so long as the cross sections are not regularized properly]:
\begin{eqnarray}
\frac{d\sigma}{d\cos\Theta}\,[q \bar{q} \to \gamma\gamma]\, &=&
     \frac{2 \pi \alpha^2 }{3 s}\, Q_q^4 \,
     \frac{1}{\sin^2\Theta} \,
   \left[ 1 + \cos^2\Theta \right]                                  \\[2mm]
\frac{d\sigma}{d\cos\Theta}\, [q \bar{q} \to Z\gamma]\, &=&
     \frac{2 \pi \alpha^2}{3 s}\, Q_q^2 [v_q^2+a_q^2] \,
     [1-M^2_Z/s]\,
     \frac{1}{\sin^2\Theta} \,
   \left[ 1 + \cos^2\Theta + \frac{4 M^2_Z s}{(s-M^2_Z)^2}\right]
\end{eqnarray}
for the invariant parton energy $\sqrt{s} = M_H$, electric and weak quark charges
denoted by $Q,v,a$ \cite{qq1,qq2}; the helicity decomposition is familiar from
electron-positron collisions, cf. appendix in Ref.$\,$\cite{Hagi}.
The parton subprocesses are peaked at small angles, regularized by the strong
interaction scale $\Lambda_{QCD}$. For the given Higgs mass, the angular
distribution of $gg \to \gamma\gamma$ \cite{gg} is surprisingly close
to the $q \bar{q}$ process, cf. Fig.~{\ref{fig:Fangle}}(b). Induced by
radiative return \cite{Min} in $Z\gamma$, and self-evident in $\gamma\gamma$,
the $\gamma$'s and the $Z$-bosons are traveling primarily along the LHC beam
axis. By restricting $|\cos\Theta|$ to less than
$\cos\Theta_{cut} \leq 1/\sqrt{2}$, the signal is reduced only modestly,
but the background strongly. \\

\noindent
{\bf 4.} For illustration of the $J^P$ sensitivity, we present a set
of rough theoretical estimates, rescaled from available experimental
data in $\gamma\gamma$ \cite{LHC_experiments1} or based on simulations
in $Z\gamma$ \cite{Schwind}. One photon with transverse energy
$\epsilon_{\text{perp}}$ in excess of 25~GeV was required for $Z\gamma$
decays, and photons in excess of 40~GeV in $\gamma\gamma$ decays. Numerically,
using $\cos\Theta = [1 - 4 \epsilon^2_{\text{perp}} M_H^2 /
(M^2_H -M_Z^2)^2]^\frac{1}{2}$
and $[1 - 4 \epsilon^2_{\text{perp}} / M^2_H ]^\frac{1}{2}$
for the corresponding polar angles in the Higgs rest frame, this is
approximately equivalent to the restrictions $|\cos\Theta| < 0.77$
and $0.55$, close to the theoretical cut $1/\sqrt{2}$ used in the
previous section. The $Z$-boson was assumed to decay leptonically
to $e,\mu$ pairs. \\

The error estimates for the $\gamma\gamma$ final states were performed
by rescaling the background event numbers of Ref.~\cite{LHC_experiments1}
in energy, using the theoretical energy dependence of $q \bar{q} \to
\gamma\gamma$ [ignoring mis-identification from jets in this simplified
theoretical estimate], and by raising the luminosity to 100 fb$^{-1}$, leaving
us with $4.6k$ [$146k$] events after cuts for signal [background],
in rough agreement, within a factor two with Ref.~\cite{duhrssen}, after
inserting the proper $K$-factor. The theoretical
angular distribution of the photons in the center range left by the cuts
was used as well according to Fig.$\,${\ref{fig:Fangle}}. We have adopted
the experimental efficiency of 40\% and a resolution of $\pm 2$ GeV.
Experimental refinements such as smearing effects etc. have not been
considered in this coarse theoretical picture. \\

In the same way as Ref.$\,$\cite{Schwind} we determined the background event
number from the cross section of the $q \bar{q} \to Z\gamma$ process
which dominates the background compared with $gg$ collisions and
the Drell-Yan process including final-state photon radiation.
Similarly to the parameters in Ref.$\,$\cite{Schwind} we have adopted
the values of 3 GeV for the mass resolution and 0.13 for the
efficiency{\footnote{These parameters were extracted by means of
PYTHIA \cite{PYTH} and ACERDET \cite{ACER}, with support by D.Zerwas
gratefully acknowledged. The transverse momentum of the photon was
chosen in excess of 25~GeV. In $Z$ decays to leptons, electrons or muons
were reconstructed with transverse energy/momentum of more than 25~GeV,
separated from the photon by $\Delta R \geq 0.7$, where $\Delta R$ is the
square-root of the sum of the squares of the azimuthal angle and pseudo-rapidity
differences. The invariant mass of the lepton pair was chosen compatible with
the mass of the $Z$ boson within 5~GeV. The reconstructed mass of the Higgs
boson was allowed in a window of 3~GeV centered on its nominal mass.
The second window can be chosen more restrictive as the Higgs boson width
is negligible compared to the experimental resolution, whereas for the $Z$
boson this is not the case. The resulting efficiency of 0.13 is close
to the value reported in
Ref.$\,$\cite{Schwind} when interpolating slightly
different parameters. }}.
The production cross section of the signal was recalculated theoretically
by including the large $K$-value in Higgs-boson production $pp \to H$. Finally,
$1.2k$ [$51k$] signal [background] events were predicted
for a luminosity of 3 ab$^{-1}$. \\

These theoretical estimates of rates and errors, not including detailed
experimental refinements, should serve only as a rough illustration of
theoretical expectations for Higgs spin analyses in radiative decays. \\

\begin{figure}[t]
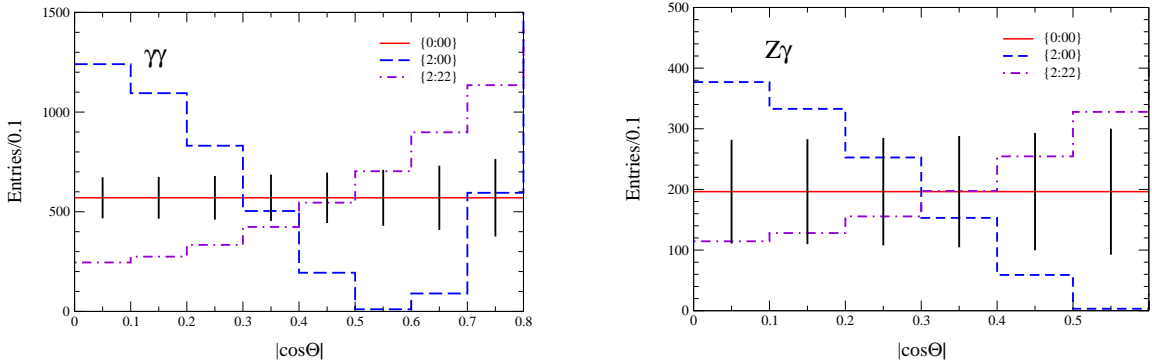

\epsfig{file=gg_sim.eps,width=7cm} \hspace{10mm}
\epsfig{file=Zg_sim.eps,width=7cm}
\caption{\it{Left: The angular distribution of the decay axis for spin-0
         Higgs-boson decays $H \to \gamma\gamma$; the spin-0 expectation
         is compared with the scalar and tensor-type decay distributions
         of a hypothetical spin-2 particle [theoretical errors only for
         the LHC luminosity = 100 fb$^{-1}$, event numbers scaled up
         from \cite{LHC_experiments1}]; Right: The same distributions for
         $H \to Z\gamma$ decays [LH-LHC luminosity = 3~ab$^{-1}$].}}
\label{fig:F.dat}
\end{figure}
In the framework defined above, the results for the signals and the size
of the roughly expected backgrounds are shown in Fig.$\,${\ref{fig:F.dat}}
for $H \to \gamma\gamma$ and $H \to Z\gamma$ in the theoretically cut
$\cos\Theta$ range. The solid lines are the [ideal] signals with very
high statistics, the error bars, based on the event numbers defined above,
represent the theoretical estimates of background fluctuations $\sqrt{B}$,
contaminating the angular distributions of the signals $S \;[\,\ll\! B]$.
These distributions are compared with the expected decay characteristics of
a hypothetical spin-2 particle, even/odd parity, derived for the
representative angular tensor-type and scalar-type distributions in
Eqs.$\,$(\ref{angles}). Note that the two distributions are mutually
complementary to each other. The first $|\cos\Theta|$ moments
[normalized to zeroth moments] are in the ratio $(0.52 \pm 0.06)/(0.27 \pm 0.05)$
for the tensor/scalar assignment in $\gamma\gamma$ final states, and
$(0.36 \pm 0.10)/(0.18 \pm 0.08)$ in $Z\gamma$ final states. Even if added up,
the resulting flattish behavior is fractured by the non-zero ${\mathcal{D}}_{20}
= {\mathcal{D}}_{02}$ contributions.  \\

The small leptonic $Z$ branching ratio together with the increased background
cross section render experimental $Z\gamma$ analyses much more demanding
than $\gamma\gamma$ analyses, and a large increase of luminosity is required. \\

Studying experimentally the $\gamma\gamma$ and $Z\gamma$ processes
outside the Higgs mass window will yield a good understanding of the background
shapes and normalizations. At the expense of a $\sqrt{2}$ increase of errors
one could define a control region with a lower
Higgs-type mass window to determine the shapes and use Monte Carlos
to extrapolate from the control region to the signal region. \\

\noindent
{\bf 5.} Summary: Dynamical characteristics of the Higgs boson in the Standard
Model are particularly difficult to analyze experimentally in the mass region
around 126~GeV since the overwhelming decays are $b$ decays. In this report we
have analyzed the theoretical potential of two decay modes, $H \to \gamma\gamma$
decays and $H \to Z\gamma$ decays [the latter statistically more remote],
to measure the spin of the Higgs boson. General helicity analyses prove
the sensitivity of both decay modes to zero-spin of the Higgs boson,
demonstrated by confronting $J = 0$ zero-spin for illustration to $J = 2^\pm$,
i.e. spin = 2 and even/odd parity.  \\[2mm]

\noindent
{\bf Acknowledgements.} The work of SYC was supported by Basic Science Research
Program through the National Research Foundation (NRF) funded by the Ministry
of Education, Science and Technology (2012-0002746). The authors would like
to thank M.~Schumacher and D.~Zerwas for useful experimental support.
A correspondence on the selection rules in Table~\ref{tab:SR} with
S.~Weinberg is greatly acknowledged.

\end{document}